\documentclass[reprint,amsmath,amssymb,aps,prl]{revtex4-1}

\usepackage{graphicx}
\usepackage{dcolumn}
\usepackage{bm}
\usepackage{color}

\begin{document}


\title{Magnetoelastic coupling in URu$_2$Si$_2$: probing multipolar correlations \\ in the hidden order state}%

\author{Mark Wartenbe}
 \affiliation{NHMFL, Florida State University, Tallahassee, FL 32310, USA}

\author{Eric D. Bauer}
\affiliation{MPA-CMMS, MS K764, Los Alamos National Laboratory, Los Alamos, NM 87545, USA}

\author{Carolina Corval\'{a}n Moya}
\affiliation{MPA-MAG, MS E536, Los Alamos National Laboratory, Los Alamos, NM 87545, USA}

\author{Ryan E. Baumbach}
 \affiliation{NHMFL, Florida State University, Tallahassee, FL 32310, USA}

\author{Neil Harrison}
\affiliation{MPA-MAG, MS E536, Los Alamos National Laboratory, Los Alamos, NM 87545, USA}

\author{Arkady Shekhter}
 \affiliation{NHMFL, Florida State University, Tallahassee, FL 32310, USA}

\author{Ross D. McDonald}
\affiliation{MPA-MAG, MS E536, Los Alamos National Laboratory, Los Alamos, NM 87545, USA}

\author{Gregory S. Boebinger}
 \affiliation{NHMFL, Florida State University, Tallahassee, FL 32310, USA}

\author{Myron B. Salamon}
\email{salamon@utdallas.edu}
\affiliation{MPA-MAG, MS E536, Los Alamos National Laboratory, Los Alamos, NM 87545, USA}

\author{Marcelo Jaime }
\email{email: mjaime@lanl.gov, corresponding author}
\affiliation{MPA-MAG, MS E536, Los Alamos National Laboratory, Los Alamos, NM 87545, USA}

\date{\today}

\begin{abstract}
Time reversal symmetry and magnetoelastic correlations are probed by means of high-resolution volume dilatometry in URu$_2$Si$_2$ at cryogenic temperatures, and magnetic fields sufficient to suppress the hidden order state at $H_{HO}(T$ = 0.66 K$) \simeq 35$ T. We report a significant magnetoelastic volume expansion at and above $H_{HO}(T)$, and even above $T_{HO}$, possibly a consequence of field-induced \em{f}\rm-electron localization. We investigate in detail the magnetostriction and magnetization as the temperature is reduced across two decades in temperature from 30 K where the system is paramagnetic, to 0.5 K in the realm of the hidden order state. We find a dominant quadratic-in-field dependence $\Delta L/L \propto H^2$, a result consistent with a state that is symmetric under time reversal. The data shows, however, an incipient yet unmistakable asymptotic approach to linear ($\Delta L/L \propto 1-H/H_0$) for 15 T $<H<H_{HO}(0.66$ K) $\sim$ 40 T at the lowest temperatures. We discuss these results in the framework of a Ginzburg-Landau formalism that proposes a complex order parameter for the HO phase to model the (H,T,p) phase diagram.

\end{abstract}

\pacs{Valid PACS appear here}
\maketitle


Despite decades of research, URu$_2$Si$_2$ remains among the most fascinating and puzzling of correlated electron systems \cite{mydosh14}. At the focus of the puzzle is the appearance of an ordered phase, heralded by a large specific heat anomaly at 17 K. The nature of the order underlying this phase remains ambiguous, hence the term “hidden order” (HO) phase. There have been numerous theoretical attempts to close the loop and many experimental probes to distinguish among them. These approaches can be divided into two classes: one which assumes that the material is primarily a band metal, with the U electrons fully hybridized with band electrons derived from Ru and Si orbitals, the other assuming that the U-atom configuration is 5$f^2$ with the HO phase evolving from either singlet or doublet crystal field ground states. There is ample experimental evidence to support each approach. Both band and CEF approaches can explain the strong anisotropy of magnetic properties, but each has difficulty determining the HO order parameter. Many such order parameters (lower rank electric multipolar, magnetic octopolar) can be eliminated as candidates \cite{yanagisawa18}.

There is some evidence in the literature for breaking of time reversal symmetry (NMR \cite{bernal06, takagi12}, μSR \cite{amato04} and Kerr rotation \cite{schemm15}). One objective of this work is to carry the study of magnetostriction to high magnetic fields in a search for broken time-reversal symmetry. We are further motivated by the recent observation by Kung et al. \cite{kung15, kung16} of a sharp feature in Raman scattering with $A_{2g}$ symmetry which appears below the HO transition $T_{HO}$ = 17.5 K. That feature has been tracked by Buhot \cite{buhot14} as a function of applied field and found to decrease in strength toward the HO critical field $H_{HO}$ = 34 T. The Raman feature was demonstrated to be consistent with the electric-hexadecapole order parameter proposed by Haule and Kotliar \cite{haule10} and by Kusunose and Harima \cite{kusunose11}. The detailed interaction between the proposed hexadecapole order and thermal expansion was calculated in mean-field theory by Haule and Kotliar. We extend that model to treat magnetostriction and demonstrate that it explains, in detail, the asymptotic approach toward linear-in-field magnetostriction reported here. We have also measured the magnetization at various low temperatures to 34 T and find it to be strikingly linear in field with no apparent correlation to the magnetostriction. Our pulsed field measurements reproduce the rich cascade of low-temperature, high-field phases [7-11] seen above $H_{HO}$. None of the results reported here, including the asymptotic approach to linear-in-field magnetostriction, supports broken time-reversal symmetry in the HO phase. Indeed, the proposed hexadecapole ordering preserves that symmetry.

We use an optical fiber Bragg grating-based dilatometry technique described before \cite{jaime17b}. Single crystal samples of URu$_2$Si$_2$ were grown by the Czochralski technique, described elsewhere \cite{schemm15}, and oriented by Laue diffraction in backscattering geometry. Bar-shaped samples were cut of approximate dimensions 2$\times$0.5$\times$0.5 mm$^3$ with the longest dimension along the principal crystallographic axis $a$ or $c$. Axial strain is obtained when the fiber is mounted parallel to the applied field, transverse strain can be measured when the sample space in the magnet permits bending the Bragg-grating-furnished end of the fiber perpendicular to the magnetic field without losing the internal reflection condition. A resolution $\Delta L/L \sim$ 1 part per million (ppm) is achieved in pulsed fields and $\sim$ 0.03 ppm in continuous fields \cite{daou10, jaime12, radtke15, jaime17b}. Complementary magnetization measurements were accomplished using a sample extraction method, where the ultimate resolution benefits from pulsed magnetic fields.  Measurements were carried out in 60 T and 65 T pulsed, and in 35 T continuous, electromagnets at the NHMFL.

Fig. 1 shows a combination of field (main panels) and temperature (insets) dependent dilatometry data. The insets in panels (a-c) show zero field (H=0) strain $\Delta$L/L vs. T where ($\Delta$L/L)$_{H=0}$ = [L(T,0)-L(25 K,0)]/L(25 K,0) in units of ppm. In panel (a) inset, we see that the $a$-axis [010] shrinks with a decrease in temperature, displaying an anomaly upon entering the HO phase (see arrow). The $c$-axis [001] in panel (b) inset, however, expands as expected from the Poisson rule. The volume effect is calculated as $\Delta$V/V = $\Delta c/c$+2$\Delta a/a$ and displayed in panel (c) inset. The computation is justified by the fact that the material is tetragonal. The coefficient of volumetric thermal expansion $\beta$(T)$ = \partial({\Delta a/a})/\partial{T}= \alpha_c$(T)$ + 2\alpha_a$(T) (not shown) is in close agreement with earlier data \cite{devisser86}. Main panels show the magnetostriction at constant temperature, computed as ($\Delta$L/L)$_{T=const}$ = [L(T,H)-L(T,0)]/L(T,0), vs. magnetic field $H \parallel$ [001] at various temperatures T with the $a$-axis, panel (a), expanding and the $c$-axis, panel (b), contracting in fields large enough to suppress the HO. An important lattice effect is observed even at T = 20 K, above T$_{HO}$=17 K. We calculate the magneto-volume change in field in panel (c). At base temperature T = 1.3 K we see very small changes at low fields, until reaching H$_{HO}$ $\simeq$ 35 T when the HO is suppressed. At this field we see a large increase in volume, significantly bigger than observed upon cooling at T$_{HO}$ in the thermal expansion data, in a series of transitions to high field phases \cite{jaime02,harrison03,kim03,kim04}. We finally see a break in slope at 39 T and a continued increase in volume  in the polarized metal regime. A similar trend is observed at 4K, which is also in the HO regime, but only a smooth evolution at 20 K.

\begin{figure}[t]
\includegraphics[scale=0.44]{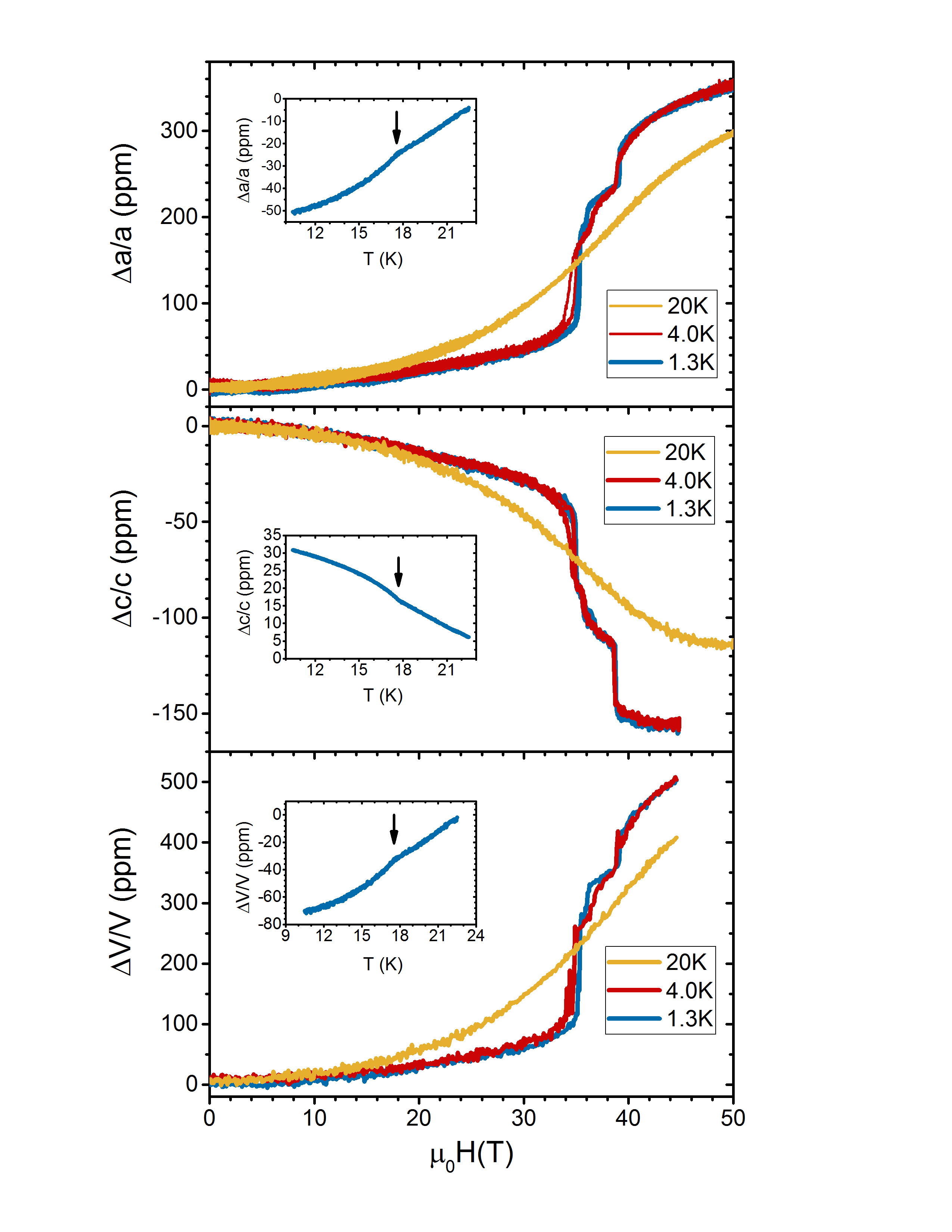}
 \caption{Panel (a) shows transverse magnetostriction ($\Delta$a/a)$_{T=const}$ measured along  [010] ($a$-axis), while panel (b) shows axial ($\Delta$c/c)$_{T=const}$ along [001] ($c$-axis). The field is applied along [001] in both cases, at various temperatures between 1.3 K and 20 K.  The insets in panels (a) and (b) show strain ($\Delta$L/L)$_{H=0}$ vs temperature.  Panel (c) shows the calculated volume dilation ($\Delta$V/V)$_{T=const}$. The inset in panel (c) shows the calculated ($\Delta$V/V)$_{H = 0}$.}
\end{figure}

The field at which the volume magnetostriction curves taken at different temperatures cross, running into each other, marks the point where the coefficient of volumetric thermal expansion ($\alpha_V$) changes sign. Note that volume magnetostriction curves are computed relative to zero field. When a 40ppm contraction between 18K and 4K is taken into account the crossing point moves slightly towards higher fields $\simeq$ 36-37T, in the realms of phase III as discussed below. Also, the crossing point is impacted by a small phonon contribution. The existence of a crossing point, however, is not impacted. Because $\alpha_V \propto V^{-1} (\partial$S/$\partial$B)$_T$ the sign change indicates accumulation of entropy that precedes a quantum critical endpoint (QCEP) \cite{garst2005}. These results confirm linear expansion data by Correa et al. \cite{correa12}. This putative QCEP was never found in URu$_2$Si$_2$ and is presumed avoided by the presence of so-called phase III \cite{harrison03,kim03}, resembling the case of Sr$_3$Ru$_2$O$_7$ \cite{gegen2006}. The large ($>$500 ppm) magneto-volume expansion observed overall, whether URu$_2$Si$_2$ is in the PM or HO state in high magnetic fields, is highly suggestive of $f$-electron localization-driven effect. Indeed, in the Kondo or partially-arrested Kondo state that develops out of a doubly degenerate ground state at low temperatures \cite{haule09}, the increasing magnetic field makes the transfer of atomic $f$-electron weight to conduction electrons less favorable. Direct exchange interaction among them, hence, likely results in a swollen unit cell volume. The more abrupt changes observed as the HO phase is suppressed point to a very strong anticorrelation with the degree of localization, $i.e$ the HO benefits from a certain degree of itinerancy and protects it but perishes at high fields, where localization is favored. As discussed below an alternative explanation includes the suppression of dipole moments on the U ions due to the onset of multipolar order for T $< T_{HO}$.

Of special interest in our data is a region in magnetic fields 15T $< \mu_0$H $ <$ 30 T where the magnetostriction appears to be remarkably linear in field, also observed by Correa at al. \cite{correa12} but never before followed in detail to the zero field limit. Fig. 2 illustrates the evolution of the $c$-axis magnetostriction vs $H$ (left panel), and vs $H^2$ (right panel), with temperature in the HO phase. At high temperatures, well above $T_{HO}$, the magnetostriction follows very closely a quadratic field dependence. The low-temperature data (0.66 K, 1.3 K, 4 K) deviate from H$^2$ as the field increases. Indeed, the lowest temperature data appear to follow a simple hyperbolic function $\Delta c/c = 1-\sqrt{1+(\rm{H}/H\it_0)^{2}}$ which is asymptotic to 1-$|$H/H$\it_0|$. This form is, of course, even under time reversal. Intermediate temperatures (7.5 K and 12.5 K) appear to transition between the two regimes. Deviation of the pulsed-field (blue) from steady-field (orange) data is a consequence of the magnetocaloric effect \cite{jaime02}.

\begin{figure}[h]
\includegraphics[scale=0.60]{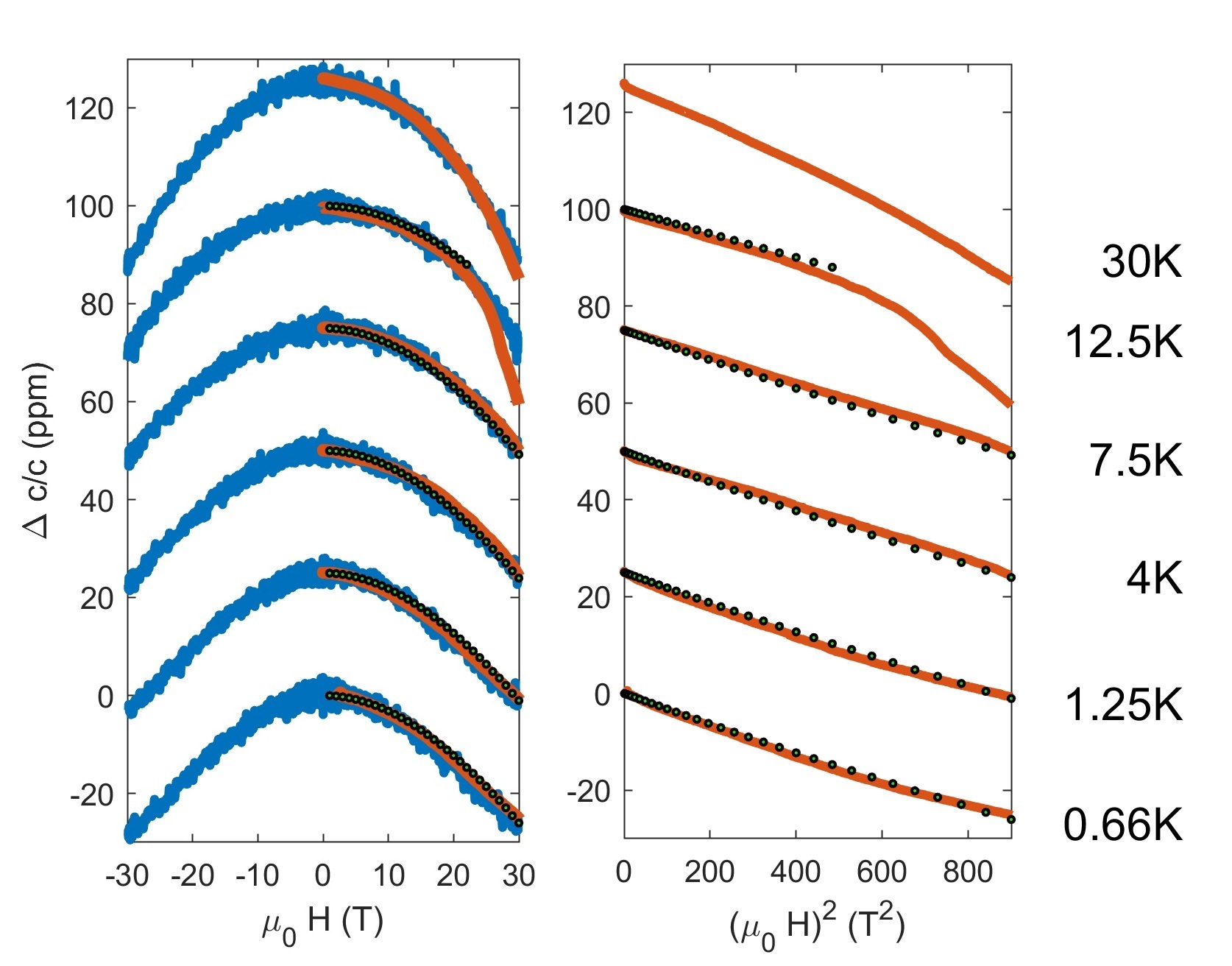}
 \caption{Waterfall plots of axial magnetostriction $\Delta$c/c versus magnetic field for linear (H) and quadratic (H$^2$) scales. Data taken in pulsed fields is displayed in blue, steady magnetic fields in red, values computed with Eq. (6) are green circles. The high-temperature magnetostriction follows a clear H$^2$ dependence. The low-temperature data (T$\ll T_{HO}$) follows a hyperbolic like behavior that becomes more pronounced at 0.66 K. No remanence or hysteresis, such as observed in piezomagnetic UO$_2$ \cite{jaime17a} is observed near or around H=0.}
 \end{figure}

The low-temperature hyperbolic form can be traced back to a Landau-Ginzburg theory for URu$_2$Si$_2$ proposed by Haule and Kotliar \cite{haule10} to account for the effects of applied  magnetic fields and hydrostatic pressures. The theory considers the competition between antiferromagnetism and the hidden hexadecapole order phase by means of a complex order parameter of the form $\Psi_{HO}$ +\textit{i}$\Psi_{AF}$. The relevant coupling constants $J_{AF}$ and $J_{HO}$ are set by $T_{HO}=\Delta/(2$~arctanh$ (\Delta/J_{HO}))$ = 17.7 K and by $T_{AF}$ = 15.7 K for $J_{AF}$. The crystal field splitting of the low-lying singlets, determined from LDA+DMFT calculations, is set at $\Delta$ = 35 K.   The applied magnetic field is converted to temperature units via $b$ = 1.25 $\mu_B\mu_0$H/$k_B$. This set of parameters predicts a critical field separating the HO and paramagnetic phases at $\mu_0H_{HO}$ = 36 T, close to data shown below.

On entering the HO phase, a spontaneous $c$-axis strain appears, predicted to be
\begin{equation}
\epsilon_{zz} = \frac{-g_{HO}c_{13}J_{HO}\Psi_{HO}^2}{\left((c_{11}+c_{12})c_{33}+2c^2_{13}\right)^2}
\end{equation}
where $g_{HO}$ reflects the strain dependence of $J_{HO}$ and positive values of $\epsilon_{zz}$ correspond to compression.  The order parameter in the  limit where a Landau-Ginzburg expansion is valid is given by 
\begin{equation}
\Psi^2_{HO}(b,T)=\frac{J_{HO}-2a(b,T)}{4u(T)}
\end{equation}
where $a(b,T))=(\Delta\lambda(b)/2)\coth(\beta\lambda(b)\Delta/2)$ ,  $u(T)=(\Delta/8)(\sinh(\beta\Delta)-\beta\Delta)(\cosh^2(\beta\Delta/2))/\sinh^2(\beta\Delta/2)$ and
\begin{equation}
\lambda(b) = \sqrt{1+\left(\frac{2b}{\Delta}\right)^2\left( \frac{J_{HO}}{J_{HO}+J_{AF}} \right)^2}
\end{equation}

Note that, at low temperatures, the hyperbolic cotangent in the expression for $a(b,T)$ approaches unity and the magnetostriction then follows the simple hyperbolic law, with $H_0$ identified as
\begin{equation}
\mu_0H_o = \left(\frac{k_B\Delta}{2.5\mu_B}\right)\left(\frac{J_{HO}+J_{AF}}{J_{HO}}\right)=40 \rm \hspace{1mm}T
\end{equation} 

The hidden order state is paramagnetic and therefore is not strongly coupled to the magnetic field. The AFM order $\Psi_{AF}$, which vanishes at zero field in the HO phase, acquires a non-zero value upon the  application of a field, reflected by the appearance of $J_{AF}$ in $\lambda(b)$. The above expression for the order parameter vanishes at $T_{HO}$ in agreement with the prediction of Haule and Kotliar. 

To compare with the experiment, we calculate
\begin{equation}
\Psi^2_{HO}(b,T) - \Psi^2_{HO}(0,T) = \frac{a(0,T)-a(b,T)}{2u(T)}
\end{equation}  
and treat  $g_{HO}$  as an adjustable parameter to fit the field dependence of $\epsilon_{zz}(B,T)$. Combining the elastic constants reported by B. Wolf, et al. \cite{wolf94}, with $J_{HO} = 3.9 \times 10^7 erg/cm^3$ (from T$_{HO}$ and the U-atom density), we find the magnetostrictive strain to be
\begin{equation}
\Delta c/c=8.4 g_{HO} (\Psi^2_{HO}(b,T) - \Psi^2_{HO}(0,T))\rm \hspace{1mm}ppm.
\end{equation} 
 Clearly, the onset of $\Psi^2_{HO} > 0$ first increases the $c$-axis lattice parameter on cooling, and then causes magnetostriction as $\Psi^2_{HO} \rightarrow 0$ in high fields. With the calculated $\Psi^2_{HO}(0)$ = 0.32 and the measured high-field $\Delta c/c$ = -26.1 ppm, the coupling strength is  $g_{HO}$ $\sim$ 9.7. The calculated points are shown as green solid circles in  Fig. 3, following the experimental curves closely. This value is consistent with the $c$-axis thermal expansion between 20 K and base temperature ($\sim$ 20 ppm), as seen in Fig. 1. The hidden-order exchange energy increases under hydrostatic pressure as $J_{HO}(P)=J_{HO}(1+g_{HO}P/c_b)$, where c$_b$ is the bulk modulus.  A similar computation for the measured high-field $\Delta a/a$ = 42.7 ppm, with -$c_{13}$ replaced by $c_{33}$ in Eq. 1,  results in a comparable coupling constant $g_{HO}$ $\sim$ 9.1. Our values predict a consequent increase in $T_{HO}$ of 2.3 K and 2.0 K at 1 GPa (using $c$- and $a$-axis magnetostriction data respectively), which compares well with a reported value of 1.7 K/GPa \cite{kambe13}.

One important consequence of broken time-reversal symmetry in a magnetic system is that linear magnetostriction $\Delta L/L \propto H$ is allowed \cite{jaime17a}. Another consequence is piezomagnetism, \it{i.e.}\rm \hspace{0.1mm} a magnetization that is proportional to magnetostriction. We carried out magnetization measurements in pulsed magnetic fields, at base temperatures, to probe piezomagnetism. Results of these measurements, in overall agreement with earlier results obtained at higher temperatures \cite{harrison03,kuwahara13,scheerer12}, are displayed in Fig. 3. The observed magnetization in the HO phase is close to linearly proportional to the magnetic field, with no evidence of H$^2$ dependence as observed in the magnetostriction, yet it departs from linear behavior for H$>$30T as H approaches H$_{HO}$. Turning the argument around it is straighforward to show that a linear-in-field magnetization M = $\chi$H in combination with the expression $\partial(\Delta c/c)/\partial H = -\partial M/\partial \sigma_c$, from Maxwell's relations where $\sigma_c$ is the uniaxial stress along the $c$-axis, lead to a magnetostriction $\Delta c/c = -1/2 \partial\chi/\partial\sigma_c H^2$. The M  $\propto$ H finding amounts, again, to the absence of evidence for broken time-reversal symmetry in URu$_2$Si$_2$. Materials that exhibit piezomagnetism also show strain hysteresis loops \cite{jaime17a}. Our attempts to detect such loops yielded no indication of a component of the strain that is linear in magnetic field, \em{i.e.}\rm~ no spontaneous moment, either in zero-field- or in field-cooled conditions. Shubnikov-de Haas quantum oscillations data from Altarawneh et al. \cite{altarawneh12} show a change in the Fermi surface of URu$_2$Si$_2$  at $\mu_0$H = 17 T. Such  Fermi surface effect, while not ruled out for the magnetostriction, would impact magnetization as well yet  we do not observe such correlation.

\begin{figure}[h]
\includegraphics[scale=0.16]{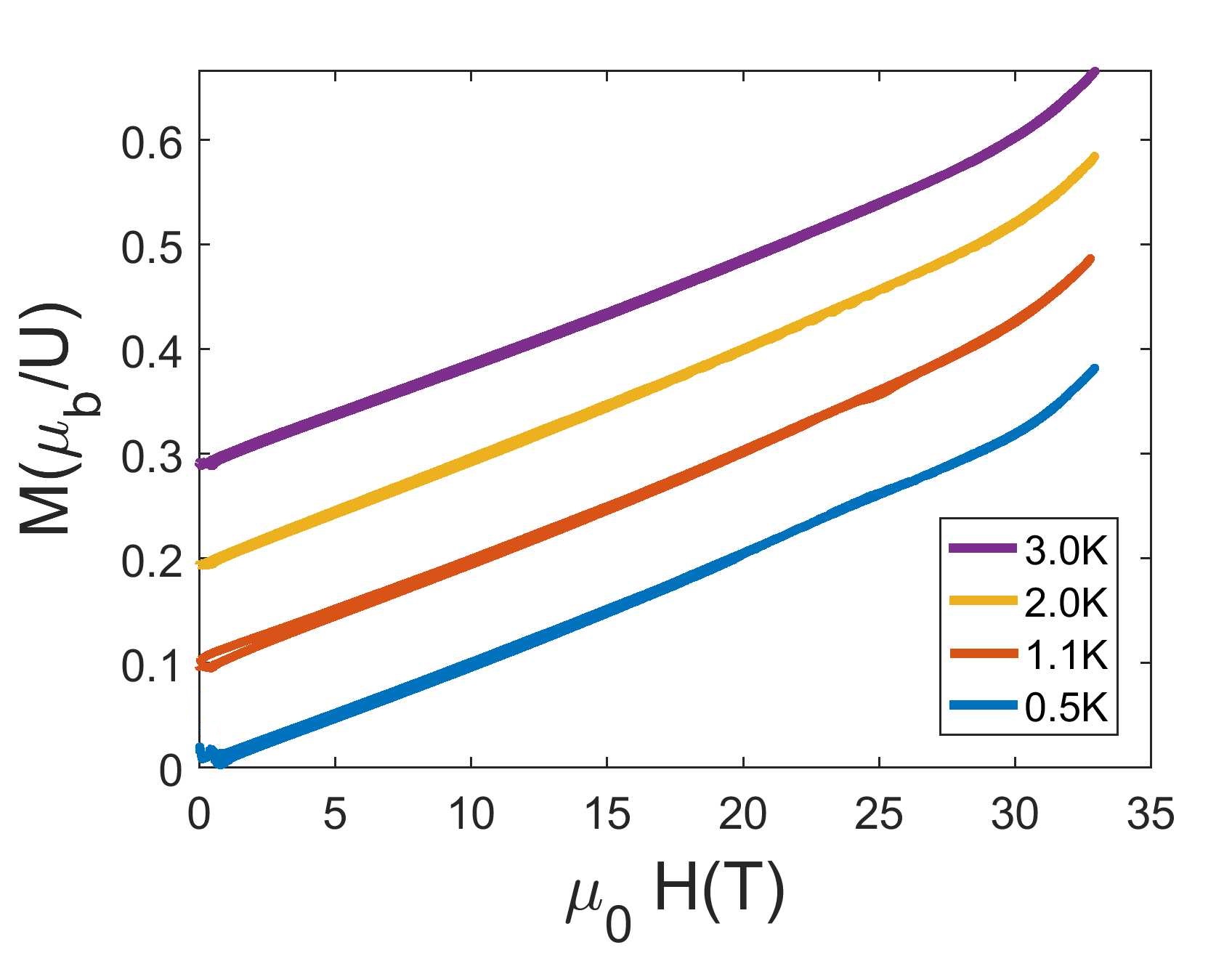}
 \caption{Waterfall plot of magnetization versus field (including field upsweep as well as down sweep) for low temperatures, showing linear behavior in the HO state, and no remanence.} 
\end{figure}


Fig. 4 shows the axial $\Delta$c/c vs magnetic field (shifted vertically for clarity) up to 42 T measured at 0.66 K, 1.3 K, and 4 K.  In agreement with the calculation for $\mu_{0}$H$_{HO}$ above, we see the suppression of the HO state at $\mu_{0}$H = 35 T (open color circles), a hysteretic region that widens as the temperature drops defined by back-dotted circles, and the suppression of phase III, at 39 T (solid color circles). Note that the hysteresis at $\mu_{0}$H$_{HO}$ =  35 T vanishes as the temperature drops below the $^4$He superfluid transition, it also vanishes in slower changing fields (not shown), indicating that (a) it is likely due to poor thermal link to the bath, and (b) that the transition at H$_{HO}$ is second order-like. The hysteresis seen upon entering (black-dotted circles) as well as upon suppresing phase III, on the other hand, grows as the temperature drops pointing to first order-like phase transitions. The region between phase I (HO) and phase III is distinct, \em{i.e.} \rm the magnetostriction shows a total of three plateaus as the magnetization data does \cite{sugiyama90, kuwahara13}. The high field zone of the phase diagram has been studied before with pulsed field neutron scattering probes, which uncovered an uncommensurate Bragg peak at \textbf{Q} = (0.6,0,0) emerging at $\mu_{0}$H = 35 T, T = 2K, and attributed it to a spin density wave (SDW) \cite{knafo16}. Our results show that, at T $\leq$ 1.3K the boundary into the SDW state, the strain becomes strongly hysteretic and the transition likely first order in nature. Notably, the HO and SDW phases do not seem to collide as T $\rightarrow$ 0.

\begin{figure}[h]
\includegraphics[scale=0.25]{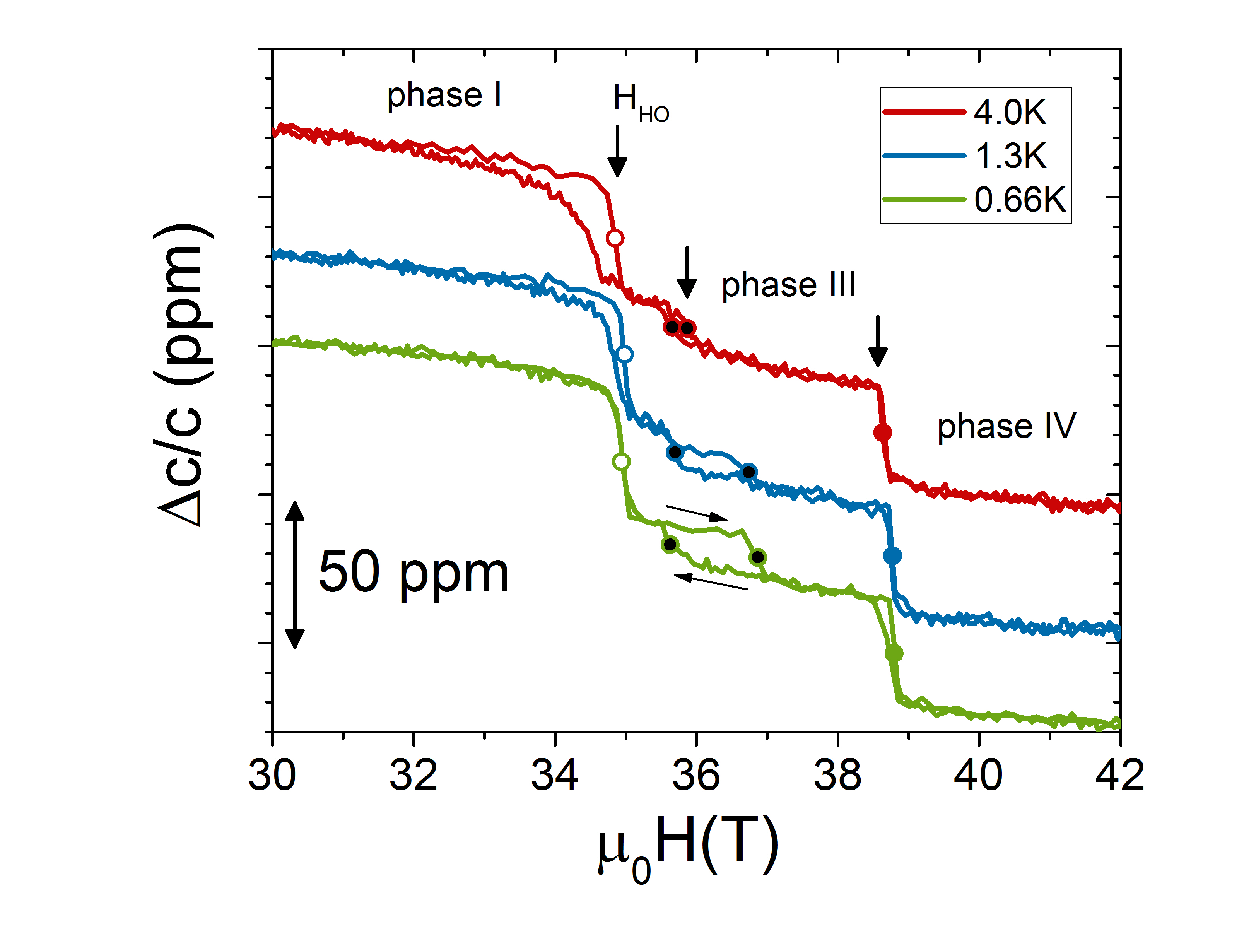}
 \caption{An expanded plot of the axial magnetostriction $\Delta$c/c(H), measured at T = 0.66 K, 1.3 K, and 4 K. Curves were shifted vertically for clarity. Here we can see the suppression of the HO phase I at H$_{HO}$ = 35T (open circles), hysteresis in the 35.5-37T range (black-dot  ted circles), and suppression of phase III at 39T (solid-color circles). Black vertical arrows indicate phase boundaries, inclined green arrows show direction of change of the magnetic field in curves measured during field up-sweep and down-sweep.}
\end{figure}


In summary, we measured the volume magnetostriction and magnetization of URu$_2$Si$_2$ to magnetic fields large enough to suppress the hidden order state. The large observed volume magnetostriction points to field-induced localization of $f$-electrons at high fields, with a clear sign change in the coefficient the thermal expansion that signals accumulation of entropy. The low field magnetostriction is predominantly proportional to $H^2$ at all temperatures, above and below $T_{HO}$, ruling out a spontaneous ordered magnetic moment by a time reversal symmetry argument. The magnetization M(H) shows a dominant linear dependence in the HO state, ruling out piezomagnetism. Consequently, no direct or indirect evidence for broken time-reversal symmetry is revealed. We found a low temperature $c$-axis magnetostriction that follows a hyperbolic function of the field, asymptotically approaching a linear dependence, in agreement with a phenomenological mean-field G-L model in the HO state that proposes a complex order parameter to model the (H,T,p) phase diagram of URu$_2$Si$_2$.  

There remain questions about the exact nature of the HO phase. While Raman scattering results reveal A$_{2g}$ symmetry pointing to hexadecapolar order \cite{kung15,kung16}, recent Ru-NQR measurements show 4-fold symmetry at the Ru and Si sites. Electric dotriacontapolar order (A$_{1u}$), with involvement of 5$f$ and 6$d$ electrons, was then proposed \cite{kambe18}. A recent ultrasonic determination of the elastic constants \cite{yanagisawa18} reported that the temperature and field dependence of $c_{11} - c_{12}$ are consistent with $A_{2g}$ symmetry, but $c_{66}$ is not. XAS and RIXS \cite{wray15} and NIXS \cite{sudermann16} studies support a doublet-ground state model, as assumed by Haule and Kotliar. We must note, however, that no pseudo-scalar order parameter, either A$_{2g}$ or A$_{1u}$-type, can by itself explain the here observed non-trivial shear response to magnetic fields. The tight connection between the HO order parameter and the local moment antiferromagnetic state in the form of a complex order parameter in the model is critical. We cannot rule out that a similar argument might be made for another HO order parameter of multipolar nature such as dotriacontapoles, and await for the development of a complementary mean-field theory that can be contrasted against our results. It is intriguing that an uncommensurare SDW state is found in phase III for H $>$ 35.5T, yet its separation from H$_{HO}$ by a first order-like field-induced transition makes its correlation with the appearance of hyperbolic-in-field magnetostriction at fields of $\simeq$15T unlikely.

Work at the NHMFL was supported by the US DOE, the State of Florida, and the National Science Foundation through cooperative grant agreements DMR 1157490 and DMR 1644779. N.H.,  M.J., and E.D.B acknowledge support by the US DOE OBES, Division of Materials Science and Engineering.


\end{document}